\documentstyle[twocolumn,aps,psfig]{revtex}

\begin{document}
\draft
\flushbottom
\twocolumn[
\hsize\textwidth\columnwidth\hsize\csname @twocolumnfalse\endcsname

\title{Metal-insulator transition in CMR materials }
\author{V. N. Smolyaninova$^{1}$, X. C. Xie$^{1,2}$, F. C. Zhang$^{1,3}$, M.
Rajeswari$^1$, R. L. Greene$^{1}$, S.~Das~Sarma$^{1}$}
\address{1. Department of Physics and Center for Superconductivity Research,\\
University of Maryland, College Park,\\
MD 20742}
\address{2. Department of Physics, Oklahoma State University, Stillwater,\\
OK 74078 }
\address{3. Department of Physics, University of Cincinnati, Cincinnati,\\
OH 45221}
\date{\today}
\maketitle
\tightenlines
\widetext
\advance\leftskip by 57pt
\advance\rightskip by 57pt

\begin{abstract}
We report on resistivity measurements in La$_{0.67}$Ca$_{0.33}$MnO$_{3}$ and Nd$_{0.7}$Sr$_{0.3}$MnO$_{3}$ thin
films in order to elucidate the underlying mechanism for the CMR behavior.
The experimental results are analyzed in terms of quantum phase transition
ideas to study the nature of the metal-insulator transition in manganese
oxides. Resistivity curves as functions of magnetization for various
temperatures show the absence of scaling behavior expected in a continuous
quantum phase transition, which leads us to conclude that the observed
metal-insulator transition is most likely a finite temperature crossover
phenomenon.


\end{abstract}

]
\narrowtext

\tightenlines


The mixed-valence perovskite manganese oxides R$_{1-x}$A$_{x}$MnO$_{3}$
(where R = La, Nd, Pr, and A = Ca, Sr, Ba, Pb) have been the materials of
intense experimental and theoretical studies over the past few years \cite
{Jin,Tokura}. These materials show colossal magnetoresistance (CMR) in
samples with $0.2<x<0.5$. In such a doping region, the resistivity exhibits
a peak at a temperature $T=T_{p}$. The system is metallic ($d\rho/dT>0$)
below $T_p$ and is insulating ($d\rho/dT<0$) above $T_p$. Under an external
magnetic field $B$, $\rho$ is strongly suppressed and the peak position $%
T_p $ shifts to a higher temperature. Thus, a huge magnetoresistance may be
produced around $T_p$ to give rise to the CMR phenomenon. It is widely
believed that the CMR behavior in these mixed-valence oxides is closely
related to their magnetic properties. This is supported by the fact that $%
T_p $ is very close to the Curie temperature $T_c$, the transition
temperature from the ferromagnetic (FM) to the paramagnetic (PM) phase.

Despite intensive investigations of the CMR phenomenon, the nature of the
metal-insulator (M-I) transition remains an open question. The
manganese-oxides are usually modeled by the double exchange Hamiltonian\cite
{Zener,Anderson,deGennes}, which describes the exchange of electrons between
neighboring Mn$^{3+}$ and Mn$^{4+}$ ions with strong on-site Hund's
coupling. As pointed out by Millis et al.\cite{Andy}, however, the double
exchange model alone does not explain the sharp change in the resistivity
near $T_{c}$ and the associated CMR phenomenon. Based on the strong
electron-phonon coupling in these materials, Millis et al.\cite{Andy}
proposed that the M-I transition involves a crossover from a high $T$
polaron dominated magnetically disordered regime to a low $T$ metallic
magnetically ordered regime. On the other hand, some authors have argued the
possible importance of the quantum localization effect (caused presumably by
the strong magnetic disorder fluctuations in the system around and above the
magnetic transition), and proposed that the M-I transition in the CMR
materials is the Anderson localization transition - a quantum phase
transition\cite{Varma,Daggoto,Houston} driven by magnetic disorder. It will
be interesting to examine the consequences of these Anderson localization
theories against experimental results.

In this paper, we report resistivity measurements in La$_{0.67}$Ca$_{0.33}$%
MnO$_{3}$ and Nd$_{0.7}$Sr$_{0.3}$MnO$_3$ thin films, 
and analyze the results to compare with the scaling
behavior expected from an Anderson localization transition. Assuming the M-I
transition in manganese oxides is of Anderson localization type, the
resistivity curves as functions of magnetization (or more precisely the
magnetic moment correlation of the neighboring Mn-ions) for different
temperatures should cross at a single point and show scaling behavior
associated with quantum criticality. Our experimental results, however,
clearly demonstrate the absence of this scaling behavior. We conclude that
the Anderson localization is not the cause of the M-I transition in the CMR
materials.

We start with a brief review of a well-known case, which exhibits the scaling
properties of the Anderson localization transition, namely the M-I
transition in thin Bi-films~\cite{Goldman}. In this case the disorder effect
is solely controlled by the thickness of the thin films, $d$. One of the
most basic scaling properties is the existence of a critical value of the
film thickness $d_c$, and a critical value of the resistivity $\rho_c$. The
resistivity is metallic for $d > d_c$, and insulating for $d < d_c$. Scaling
laws are established for physical quantities with parameters near these
critical values. Absence of these critical values would imply the absence of
scaling behavior, incompatible with the theory of the Anderson transition
which is a continuous quantum phase transition manifesting
scaling behavior.

If we assume the M-I transition in manganese oxides to be an Anderson
localization, the question then naturally arises about what would be the
physical quantity or the tuning parameter which corresponds to the layer thickness in the Bi-thin
films describing the disorder strength.  We believe that the tuning parameter in the CMR M-I transition should be
the magnetization of the system.  To make the discussion more concrete, let us
consider a model discussed in Ref. 9 to describe the possible Anderson
transition in Mn-oxides,

\begin{equation}  \label{hamil}
H_{eff}=-\sum_{ij}t^{\prime}_{ij}d_{i}^{+}d_{j}+ \sum_{i} \epsilon
_{i}d_{i}^{+}d_{i}+c.c
\end{equation}

Here the first term is the effective double-exchange Hamiltonian in which
$t^{\prime}_{ij}=t \{ \cos (\theta_i/2)\cos(\theta_j/2)
+\sin(\theta_i/2)\sin(\theta_j/2) \exp[i(\phi_i - \phi_j)] \}$, with $t$ the
hopping integral in the absence of the Hund's coupling and the polar angles
$(\theta_i,\phi_i)$ characterizing the orientation of local spin $\vec S_i$. 
The second term in Eq.(1) represents an effective on-site disorder Hamiltonian
(which should lead to the M-I Anderson transition in this model) which
includes all possible diagonal disorder terms in the system, such as the local
potential fluctuations due to the substitution of R$^{3+}$ with A$^{2+}$. Here
$\epsilon_i$ stands for random on-site energies distributed within the range
$[-W/2, W/2]$. For a given sample, the diagonal disorder, namely
$\{\epsilon_i \}$, or $W$, is fixed, but the bandwidth is controlled by the
double exchange hopping integral.  Therefore the effective strength of the disorder is
determined by $t^{\prime}$. Experimentally the disorder strength may be tuned
by introducing an external magnetic field $B$ and/or by changing the
temperature $T$. For instance, as $B$ increases, the magnetic ions tend to
align along the same direction so that the magnitude of  $t^{\prime}$, hence
the bandwidth, increases. As $T$ is lowered below $T_c$, there is 
spontaneous magnetization, which can also increase the bandwidth to reduce the
disorder strength. Note that the role of temperature in this localization
model is somewhat indirect in the sense that it only controls the disorder
strength - the usual role of finite temperature in quantum phase transition is
the introduction of a dynamical exponent $z$ which would not play an explicit
role in the discussion and analysis of the experimental data presented in this
paper.

The hopping integral $<t^{\prime}_{ij} >$ in the double exchange model
depends on the magnetic moment correlation between the neighboring Mn-ions, $%
\chi = < \vec M_i \cdot \vec M_j >$, where $<...>$ denotes the thermal
average. $\chi$ can be divided into a static part and a fluctuation part, $%
\chi = M^2 - \Delta M^2$, where $M$ is the magnetization, which can be
measured directly, and $\Delta M^2 = \chi - M^2 \geq 0$ for 
ferromagnetic interacting systems including the present case. Sufficiently
away from the magnetic transition point ($T=T_c$ and magnetic field $B=0$), the
fluctuation part can be dropped, and the bandwidth is controlled by the
magnetization.  In what follows, we first neglect the fluctuation effect,
and focus on the static part to discuss the scaling behavior. This
approximation is equivalent to the mean field approximation made in Ref. 9.
The fluctuation effect does not alter our qualitative conclusion.

The effect of off-diagonal disorder (arising, for example, from a random
$t^{\prime}$) was previously discussed by
Varma for the paramagnetic phase~\cite{Varma}. In that work, the core-spin
fluctuation was treated in the adiabatic approximation and the primary effect
of the magnetic field was argued to alter the localization length. A more
detailed calculation by Li et al.~\cite{Li}, including both random hopping
and on-site disorder, showed that random hopping
alone is not sufficient to induce Anderson localization at the Fermi level
relevant to the observed CMR phenomenon.  We believe the main effect of the
magnetic field in the high field limit is to partially
polarize the electron spins, thus to increase the electron bandwidth. Our
analyses should apply to the experimental situation reported in this paper,
where the magnetization is as high as a fraction of the saturated value.

\begin{figure}[tbp]
\centerline{
\psfig{figure=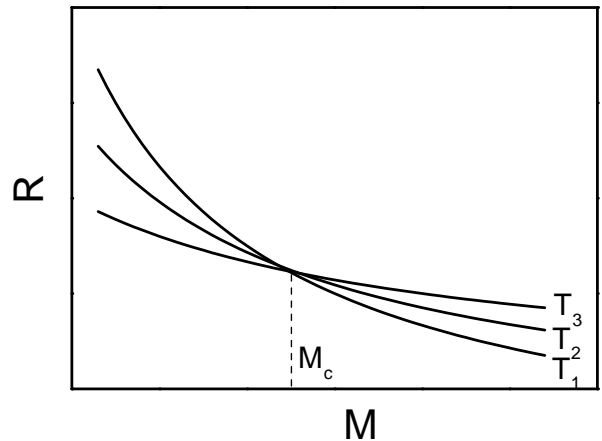,width=8.5cm,height=6.4cm,clip=}
}
\caption{The behavior of resistance $R$ shown schematically as a function of
magnetization $M$ for three different temperatures, under the assumption
that $M$ can induce the Anderson delocalization transition. The quantum
critical point is indicated by $M_c$.}
\label{fig1}
\end{figure}

Similar to the layer thickness in the Bi-thin films, we would thus expect a
critical value in magnetization, $M_c$, in the CMR materials, i.e., $M$ is
the control or the tuning parameter for the quantum phase transition. For $M
< M_c$, the system is an insulator and $\rho$ decreases with increasing
temperature. For $M > M_c$, the system is a metal and $\rho$ increases with
increasing temperature. In Fig. 1, we schematically illustrate the expected
resistivity as functions of $M$ for three temperatures $T_{1}$, $T_2$ and $%
T_3$ with $T_{1}<T_{2}<T_{3}$. All $T_i$ ($i=1,2,3$) are above the Curie
temperature $T_c$ and the peak temperature $T_p$. These different
temperature curves should cross at a single value $M_c$ if the transition is
of Anderson type. The reason for the crossing is as follows. A given
temperature gives an effective cut-off length scale. The resistivity depends
on the ratio of this length scale to the localization length. At the
critical point, the localization length diverges, thus the resistivity is
independent of temperature. Below we first present our resistivity and
magnetization measurements at various external field $B$ for different $T$.
We then analyze our results and discuss the magnetization dependence of the
resistivity for various $T$. These data will be shown to be incompatible
with the critical scaling requirement of an Anderson localization transition.


The samples used in this study are epitaxial thin films of
La$_{0.67}$Ca$_{0.33}$MnO$_{3}$ and Nd$_{0.7}$Sr$_{0.3}$MnO$_3$ grown by
pulsed laser
deposition on LaAlO$_{3}$ substrate.
The film thickness is 1200 \AA~ for La$_{0.67}$Ca$_{0.33}$MnO$_{3}$ and
2100~\AA~ for Nd$_{0.7}$Sr$_{0.3}$MnO$_3$ samples. The deposition conditions
are: laser
fluence 2 J/cm$^{2}$, substrate temperature 820$^{\circ}$C, oxygen partial
pressure in the growth ambient 400 mTorr. Following deposition, films were
cooled down to room temperatures in an oxygen ambient 400 Torr. The samples
were subjected to post-deposition thermal annealing at 850$^{\circ}$C in
oxygen for 10 hours. X-ray studies were used to ensure good structural
quality of the samples.

Resistivity was measured by a standard four-probe technique. Magnetization
was measured with a commercial SQUID magnetometer. The magnetic field was
applied parallel to the film plane in order to minimize the demagnetization
factor. The diamagnetic contribution of the substrate was measured
separately and subtracted.

The Curie temperature of the samples was determined from the temperature
dependence of magnetization at low magnetic field, and is found to be 270 K (%
$T_c$) for La$_{0.67}$Ca$_{0.33}$MnO$_{3}$ and 205 K for
Nd$_{0.7}$Sr$_{0.3}$MnO$_3$. At zero field the resistance has a peak around
$T_p\sim 275$~K for La$_{0.67}$Ca$_{0.33}$MnO$_{3}$
(Fig.2, inset) and $T_p\sim 217$~K for Nd$_{0.7}$Sr$_{0.3}$MnO$_3$, which is
close to the corresponding Curie temperatures. The peak values of
resistivity are $\sim$ 10 mOhm-cm for La$_{0.67}$Ca$_{0.33}$MnO$_{3}$ and
$\sim  145$~mOhm-cm for Nd$_{0.7}$Sr$_{0.3}$MnO$_3$  and the residual low
temperature
resistivity values are 170 $\mu$Ohm-cm for
La$_{0.67}$Ca$_{0.33}$MnO$_{3}$ and
550 $\mu$Ohm-cm for, which are
\linebreak
\begin{figure}[tbp]
\centerline{
\psfig{figure=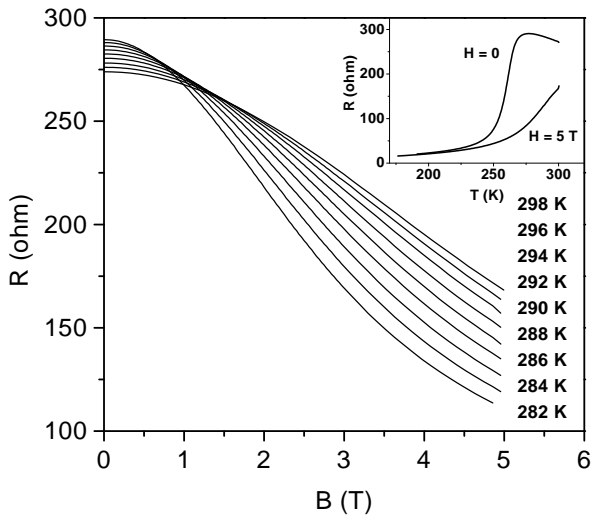,width=8.5cm,height=7.4cm,clip=}
}
\caption{Measured resistance of La$_{0.67}$Ca$_{0.33}$MnO$_3$ film versus
magnetic field for different
temperatures. Temperature dependence of the resistance at $B = 0$ and 5 T is
shown in the inset. }
\label{fig2}
\end{figure}
\begin{figure}[tbp]
\centerline{
\psfig{figure=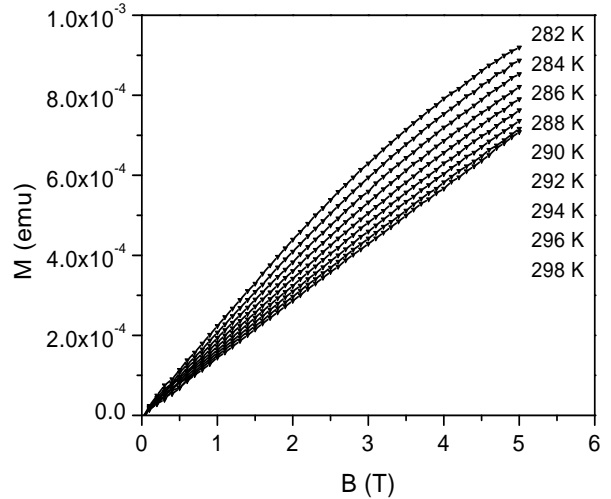,width=8.5cm,height=7.3cm,clip=}
}
\caption{Measured magnetization $M$ of La$_{0.67}$Ca$_{0.33}$MnO$_3$ film
versus magnetic field $B$ for
temperature range $T=282$ K to $298$ K.}
\label{fig3}
\end{figure}
\noindent
typical values for good quality
epitaxial films of these compositions.


We now discuss our experimental results. All the
results are from La$_{0.67}$Ca$_{0.33}$MnO$_3$ films
except those in Fig. 4(b). In Fig. 2, we show resistance $R$
as a function of magnetic field for $B=0$ to 5 T for the temperatures just
above $T_p$. At zero field, $R$ has a peak around $T_{p}\sim 275$ K, above
which the sample is an insulator. At $B=5$ T, the insulating phase has been
eliminated by the applied field, and $\partial R(B, T)/ \partial T > 0$
within the measured temperature region $275$ K$< T< 300$ K. In Fig. 3, we
show the measured magnetization $M$ as a function of magnetic field for a
temperature range between $T=282$~K and $T=298$~K.

The main result of this paper is shown in Fig. 4. In Fig. 4(a) the
resistance $R$
is plotted as a function of magnetization $M$ for several temperatures
ranging between $T=282$ K to $298$ K. These curves were obtained by
combining the data from Figs. 2 and 3. The $R(M)$ curves appear to be
approximately crossing with each other at the magnetization value about $%
3\times 10^{-4}$ emu. This crossing might appear to indicate localization
due to the reduction of the bandwidth, represented by $M$ here. 
However, there is no single crossing point for all these curves, as
shown in the inset to Fig. 4(a). Intersections of the curves occur from $%
M=2.6\times 10^{-4}$ emu to $M=3.6\times 10^{-4}$ emu. This interval is
about 15$\%$ of the studied magnetization range, which could hardly be
defined as a single point. Besides, at higher magnetization values $R(M)$
curves converge again. This result is manifestly incompatible with the
Anderson M-I transition behavior in Fig.~1 or equivalently, with the
general behavior of a continuous quantum phase transition. Therefore, we
conclude that Anderson localization is not the mechanism for the M-I
transition in La$_{1-x}$Ca$_{x}$MnO$_{3}$ thin films. In addition, we have
explicitly verified that our $R(M,T)$ data do not exhibit quantum scaling
behavior and cannot be collapsed into one effective curve by choosing
suitable localization and dynamical exponents.

\begin{figure}[tbp]
\centerline{
\psfig{figure=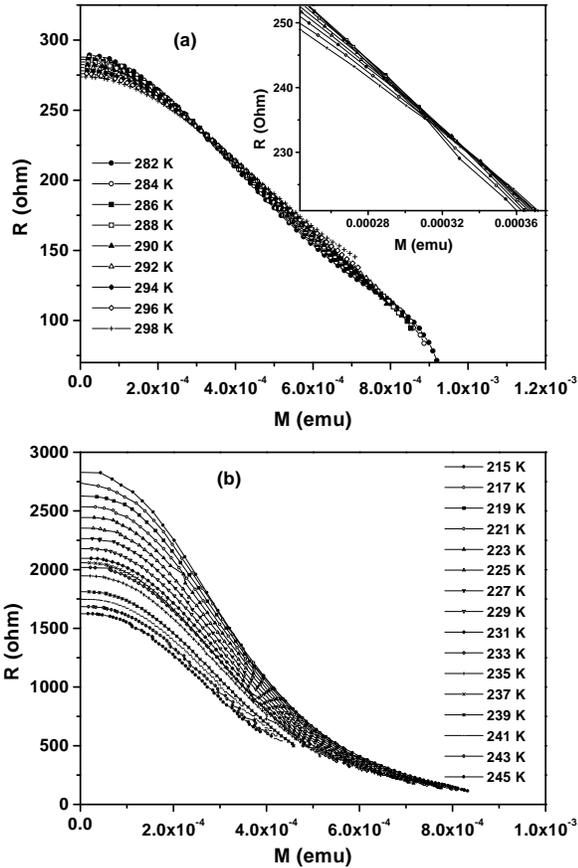,width=8.5cm,height=12.1cm,clip=}
}
\caption{(a) Resistance versus magnetization for La$_{0.67}$Ca$_{0.33}$MnO$_3$ 
for temperature range $T=282$ K to $%
298$ K. Portion of the graph near the intersections is shown enlarged in the
inset. (b) Resistance versus magnetization for Nd$_{0.7}$Sr$_{0.3}$MnO$_3$
for temperature range $T=217$ K to $T=245$ K.}
\label{fig4}
\end{figure}

To determine whether our results from the La$_{0.67}$Ca$_{0.33}$MnO$_{3}$
samples are generic, we have carried out measurements on
Nd$_{0.7}$Sr$_{0.3}$MnO$_3$ films. In Fig.~4(b) the resistance $R$
is plotted as a function of magnetization $M$ for several temperatures ranging
between 217 K to 245 K. Different curves from different temperatures
do not even cross for the Nd$_{0.7}$Sr$_{0.3}$MnO$_3$ film which is inconsistent
with the behavior of Fig. 1, expected
for a quantum phase transition.


Our main experimental conclusion, as shown in Figs.~2~-~4, is that the measured
thin film resistivity $R(M,T)$ of CMR manganite materials plotted as a
function of the measured magnetization $(M)$ at different temperatures $(T)$
does not exhibit any simple quantum criticality around the measured M-I
transition temperature $T_p$.
This is manifestly obvious from
Fig. 4 since $R(M)$ for different temperatures around $T_p$ do not
cross through a single critical magnetization value $(M_c)$ as they should if the
underlying cause is a
continuous quantum phase transition as in Anderson localization. Our analysis
has been based on the assumption that the magnetization is the appropriate
tuning parameter for the localization quantum phase transition in manganites
(i.e. the transition is
driven by magnetic fluctuations). Magnetization as the tuning parameter is
entirely reasonable for quenched disorder which we have implicitly assumed in
our analysis. If the disorder is arising entirely from temperature dependent
(intrinsic) magnetic
fluctuations, then the relevant disorder is annealed, and recent detailed
numerical work \cite{Li} shows that such intrinsic annealed disorder is
unlikely to lead to localization without additional strong quenched magnetic
disorder arising from, for
example, structural disorder. Our experimental results indicate that a
continuous quantum phase transition is unlikely to be the underlying cause for
the CMR M-I transition, and the observed phenomenon is most likely a rapid
crossover behavior at $T_p$.
We cannot, however, comment on the nature of this crossover behavior based
only on our experimental results.

One issue requiring some elaboration in the context of metal-insulator
transitions in CMR materials is the fact that phenomenologically this M-I
transition is thought to occur at a transition temperature ($T_p \sim 275$ K,
217~K for the La$_{0.67}$Ca
$_{0.33}$MnO$_{3}$, Nd$_{0.7}$Sr$_{0.3}$MnO$_{3}$ materials respectively in
our experiments) with the system being ``metallic" (also ferromagnetic) for
$T<T_p$ and ``insulating" (paramagnetic) for $T>T_p$.
The effective ``metallic" and ``insulating" phases in this CMR M-I transition
are defined entirely by the temperature dependence of $\rho (T)$ with
$d\rho/dT>0$ (for $T<T_p$) defining an effective metal whereas $d\rho/dT<0$
(for $T>T_p$) defining the
effective insulator. The true M-I localization transition is a $T=0$ quantum
phase transition with the insulating phase having zero conductivity and the
metallic phase having finite conductivity. The sign of $d\rho/dT$ is not always
a good indicator
for a M-I transition. In our analyses of the data (as well as in the current
discussion on M-I transitions in CMR materials) one assumes the temperature to
be a parameter affecting the magnitudes of the physical quantities
(e.g. magnetic behavior) defining the M-I transition. It may actually be more
natural to think of the CMR
M-I transition as a temperature-induced crossover behavior, and any critical
discussion of a true M-I transition in CMR materials should await an
experimental observation
of a M-I transition at a fixed low temperature as a function of a system
parameter (e.g. disorder, magnetic impurities, sample thickness, composition).
All of the current activity on the nature of the M-I transition in CMR
materials may thus be premature unless one can experimentally induce a low
temperature transition by
varying a system parameter. In that context the most important experimental
result produced by our investigation is the finding that the resistivity
$R(M,T)$ in CMR materials
around the M-I ``transition" temperature $T_p$ {\it cannot be written simply}
as $R(M(T))$ as has been almost universally assumed in prior work
\cite{Wisconsin} on the subject. We find, as is obvious from Figs. 2-4, that
the measured resistance $R$ is {\it not} just a function of the system
magnetization $M(T)$ at that particular temperature, but is also a
function of temperature $T$
directly (i.e. $R$ at a fixed $M$, but different $T$ values, takes on
different values as can be seen in
Fig. 4). Thus $R$ is a two parameter function $R(M,T)$ with $M(T)$ depending
also on $T$. While the direct temperature dependence of $R$ is not extremely
strong,  it is clear that $R$ cannot be expressed as a
simple one parameter
function $R(M(T))$. We believe that this finding should have important
implications for the CMR phenomena which far transcends the specific issue
of whether the observed CMR M-I transition is a continuous quantum phase
transition or a crossover behavior. We emphasize that the non-existence of a
critical $M_c$ (at which $R$
for different $T$ values cross, indicating the existence of a single M-I
transition point), which is the main factual finding of our work, implies that
there is {\it no} magnetization independent M-I transition in CMR materials
induced only by
temperature - the measured resistance {\it always} depend on both $M$ and $T$.
We also note that our measured resistance can be approximately fitted by an
exponential function in $M/M_{sat}$, but such fits are manifestly approximate
since the measured
resistance always depends on both $M$ and $T$ {\it independently}.

One may question our choice of the
magnetization as the control parameter in driving the M-I transition in
contrast to, for example, the applied magnetic field, which superficially may
appear to be the tuning parameter for the Anderson localization. We believe
the appropriate tuning parameter to be the magnetization (at least with the
double exchange Hamiltonian defined in Eq. (1)), since it
determines the effective hopping integral $t^{\prime }$, and hence the
disorder strength in the Hamiltonian. We have of course studied the
resistivity as a function of the magnetic field in the temperature range $%
T=282$ K to $298$~K, as shown in Fig. 2. No single transition point and/or
quantum scaling can be defined from the magnetic field study in Fig. 2,
leading to the same conclusion about the non-existence of a continuous M-I
transition. A more appropriate quantity to characterize the disorder
strength in the manganese oxides would perhaps be the magnetic moment
correlation $\chi $ of the neighboring Mn-ions, which is difficult to
measure directly. A quantitative experimental study of resistivity as a
function of $\chi $ for various $T$ would be very difficult. We can,
however, make a general statement that a measurement of $R(\chi ,T)$ is
unlikely to exhibit quantum critical scaling because our $R(M,T)$ data
manifest non-scaling behavior in Figs. 2-4. We believe that our measured $%
R(T,M)$ behavior is exhibiting the intrinsic metal-insulator crossover in
the system, and there is no continuous metal-insulator phase transition in
the problem. 

In conclusion, we have carried out resistivity measurements in La$_{1-x}$Ca$%
_{x}$MnO$_{3}$ and Nd$_{1-x}$Sr$_x$MnO$_3$ thin films 
to study the possible Anderson
metal-insulator transition. An external magnetic field is applied to induce
the paramagnetic to ferromagnetic transition. As a function of
magnetization, the resistivity curves for different temperatures are found 
{\bf not} to cross at a single point, establishing the non-existence of a
quantum critical point. This result is incompatible with theoretical
expectations from Anderson metal-insulator transition. Thus, we conclude
that the Anderson localization is not the cause of the metal-insulator
transition in La$_{1-x}$Ca$_{x}$MnO$_{3}$ thin films. The precise nature of
the metal-insulator transition in CMR materials requires further
experimental and theoretical investigations. The present experiments seem to
be consistent with the picture that the transition is a crossover from a
metal to a magnetically disordered polaronic insulator\cite{Andy}.

\noindent {\bf Acknowledgment}: This work is supported by the NSF-MRSEC, NSF
and ONR at Maryland. X.C. Xie and F.C. Zhang are also supported by DOE under
the contract number DE-FG03-98ER45687.


\end{document}